\documentclass[fleqn,usenatbib]{mnras}

\usepackage{newtxtext,newtxmath}

\usepackage[T1]{fontenc}

\DeclareRobustCommand{\VAN}[3]{#2}
\let\VANthebibliography\thebibliography
\def\thebibliography{\DeclareRobustCommand{\VAN}[3]{##3}\VANthebibliography}

\usepackage{graphicx}	
\usepackage{amsmath}	

\title[UV upturn in LRGs]{The Ultraviolet Upturn in field Luminous Red Galaxies at $0.3 < z < 0.7$}

\author[R. De Propris et al.]{
R. De Propris,$^{1}$\thanks{E-mail: rodepr@utu.fi}
S. S. Ali,$^{2}$
C. Chung,$^{3}$
M. N. Bremer,$^{4}$
and S. Phillipps$^{4}$
\\
$^{1}$FINCA, University of Turku, Vesilinnantie 5, 20014, Turku, Finland\\
$^{2}$Subaru Telescope, National Astronomical Observatory of Japan,
650 North Aohoku Place, Hilo, HI, 96720, USA\\
$^{3}$Department of Astronomy and Center for Galaxy Evolution Research, Yonsei University, Seoul 03722, Republic of Korea\\
$^{4}$Astrophysics Group, School of Physics, University of Bristol, Tyndall Avenue, Bristol BS8 1TL, UK
}

\date{Accepted XXX. Received YYY; in original form ZZZ}

\pubyear{2021}

\begin{document}
\label{firstpage}
\pagerange{\pageref{firstpage}--\pageref{lastpage}}
\maketitle

\begin{abstract}
We derive the evolution of the ultraviolet upturn 
colour from a sample of field luminous red galaxies at 
$0.3 < z < 0.7$ with $-24 < M_r < -21.5$. No individual objects are securely detected, so we stack several hundred galaxies within absolute magnitude and redshift intervals. We find that the colour of the ultraviolet upturn (in observed $NUV-i$ which is approximately equivalent to the classical $FUV-V$ at the redshifts of our targets) does not change strongly with redshift to $z=0.7$. This behaviour is similar to that observed
in cluster ellipticals over this same mass range and
at similar redshifts and we speculate that the processes
involved in the origin of the UV upturn are the same.
The observations are most consistent with spectral
synthesis models containing a fraction of a helium
rich stellar population with abundances between
37\% and 42\%, although we cannot formally exclude
a contribution due to residual star formation at the
$\sim 0.5\%$ level (however, this appears unlikely 
for cluster galaxies that are believed to be more
quenched). This suggests that the ultraviolet upturn is
a primordial characteristic of early type galaxies at
all redshifts and that an unexpected nucleosynthesis
channel may lead to nearly complete chemical evolution
at early times.


\end{abstract}




\section{Introduction}

The ultraviolet upturn is the observed excess flux
at $\lambda < 3000$ \AA\ in the spectral energy distributions of early-type galaxies (ETGs) above 
the predictions of conventional stellar population
synthesis models. These models predict that such galaxies should instead 
be very dark at these wavelengths, given their high metallicities and old
ages (e.g., see reviews by \citealt{OConnell1999,Yi2008}). Typical
colours in $FUV-V$ for these ETGs (where $FUV$ is a $\sim 600$ \AA\ 
wide band centred at $1500$ \AA) are 1.5--2 mag.
bluer than those expected from a stellar population
formed at $z=4$, with solar metallicity and star-formation e-folding time ($\tau$) of 0.3 Gyr. This model 
accurately reproduces the observed colours longwards of $\sim 3000$ \AA (effectively all optical-IR colours) of ETGs to within a
few percent, even at young ages (e.g., \citealt{Bruzual2003,Conroy2009,Conroy2010}). The UV upturn therefore points to some incompleteness in our
understanding of the stellar populations of ETGs. 

There is now a broad consensus that the sources of
the UV excess light in spheroidal systems are blue horizontal branch (HB)
stars (e.g., \citealt{Greggio1990,Dorman1993,Dorman1995,Brown2004}). These objects are directly observed in
the bulge of M32 \citep{Brown2000a},
and provide (although unresolved) most of the UV flux in M31's bulge \citep{Rosenfield2012}. The $FUV$
spectra of a few nearby ellipticals taken 
by the {\it Hopkins Ultraviolet Telescope} flown
on the Space Shuttle are consistent with a 25000K blackbody (the underlying old stellar population produces negligible flux in this regime), as produced by an extended HB \citep{Brown1997}. Similarly, the UV spectral
energy distributions for ETGs in Coma and 
Abell 1689 derived by \cite{Ali2018a,Ali2018b}
are also best fit by single blackbodies of 
relatively high temperatures, with the flux and effective temperature depending on galaxy mass. The UV spectral properties of these galaxies are also broadly inconsistent with those expected from alternate sources such as young stars \citep{Vazdekis2016,Rusinol2019} or intermediate mass white dwarfs \citep{Ferguson1993,Werle2020}.

However, blue HB stars are not normally produced
in old, metal-rich stellar populations \citep[e.g.][]{Catelan2009}. For such stars to
evolve to the blue HB requires either an increased
stellar mass loss (revealing the hotter inner layers of the stellar envelopes) during the first ascent of the 
red giant branch or a He-enriched chemical composition. An increase in mass loss for metal rich 
stellar populations as proposed by \cite{Yi1997} 
is not observed in our Galaxy across more than 3 dex.
in metallicity \citep{Miglio2012,McDonald2015,Williams2018}. \cite{Han2007} and \cite{Hernandez2014} instead
propose that mass loss occurs within close binaries,
as the stellar envelopes are stripped by angular
momentum transfer as one of the stars expands 
beyond its Roche lobe. As already pointed out by
\cite{Smith2012} this cannot easily account for
the observed trend where more massive and more
metal rich galaxies have bluer UV upturns \citep{Burstein1988,Ali2018a}. Neither can this
model explain the strong radial gradients within
galaxies in the UV upturn colour and their dependence
on $Mg_2$ strength \citep{Carter2011,Jeong2012}. \cite{Petty2013} show that the observed colour gradients are consistent with an increase in the fraction of blue HB stars towards the centers of galaxies by around 25\%.
This would imply that binaries would have to be both more frequent and tighter (to explain the bluer colours) as a function of radius (and Mg$_2$ strength), a process that appears to require considerable fine-tuning.

\cite{Tantalo1996} and \cite{Chung2011,Chung2017}
argue for the presence of a minority population of
He rich stars, that can evolve to the blue HB even
at high metallicities. This occurs because the lifetimes of He rich stars
are shorter, allowing for lower mass stars to evolve
at the same time (with a correspondingly thinner stellar atmosphere over the nearly constant mass
core). Unlike other proposed sources of the UV upturn, these stars have clear counterparts in our
Galaxy, namely the multiple populations of globular clusters producing anomalously blue HBs \citep{Gratton2012,Bastian2018}.
The radial gradients observed by \cite{Carter2011,Jeong2012} and \cite{Petty2013} could then be produced by a radial gradient in He abundance, especially if this is produced by earlier stellar populations (as is observed in galactic globular clusters). Indeed, the UV upturn and the multiple populations in globular clusters could be manifestations of the same phenomenon albeit in different environments,
or the He rich stars may be produced within metal-rich globular clusters as well \citep{Goudfrooij2018,Chung2020}

Each of these explanations imply different predictions as
to the evolution and environmental dependence of the
UV upturn, and these can be used to discriminate
between them.
\cite{Ali2019} and \cite{Phillipps2020} carried out
the first comparison of UV upturn properties across
a wide range of environments at low redshifts, ranging
from clusters to groups and the more isolated field,
and they found no evidence of any strong environmental
dependence. This suggests that the UV upturn is an
intrinsic property of ETGs originating at early times
and argues against a component from residual star
formation as the UV upturn is independent of local
galaxy density, which otherwise affects star formation
properties elsewhere, including residual star formation in ETGs -- \citealt{Crossett2017}.

Most studies of the evolution of the UV upturn were originally limited to bright cluster galaxies, out to relatively
moderate redshifts \citep{Brown1998b,Brown2000b,Brown2003,Ree2007,Donahue2010,Boissier2018} although \cite{Atlee2009} studied the evolution of field ETGs out to $z=0.6$ by stacking objects. These found evidence for only moderate
to no evolution out to $z=0.6$, which does not 
provide strong evidence for either model discussed above. \cite{LeCras2016} used spectroscopic indices on stacked spectra of bright LRGs from the BOSS survey
and found that the strength of the UV upturn declines
at $z > 0.6$, nearly disappearing at $z=1$, while 
\cite{Ali2018b,Ali2018c} observed that the UV upturn
does not evolve in clusters to $z=0.55$ but then its 
colour becomes significantly redder at $z=0.7$. \cite{Ali2021} show that the $NUV-r$ colour of cluster ETGs is also consistent with the presence of an UV upturn to $z=0.6$ but this component largely disappars at $z=0.95$. This is evidence of evolution in the
UV upturn (although \citealt{Lonoce2020} claim a detection of the UV upturn in a $z=1.4$ galaxy, albeit from a single spectral line at the $2\sigma$ level) with redshift. Only models with a He-rich subpopulation
are able to reproduce the whole evolutionary trend in
the color of the UV upturn. From this,
\cite{Ali2018a,Ali2018b,Ali2018c} and \cite{Ali2021}
estimate that ETGs would have to contain a population (about $\sim 
10\%$ by mass) of stars with $Y > 0.42$ formed at 
$z > 3$, to explain the behaviour and strength of the UV upturn as a function of redshift. This level of He enrichment is similar to that observed in the more extreme multiple populations in globular clusters in our Galaxy.

Here we extend this analysis to field ETGs at $0.3 < z < 0.7$ to compare evolutionary trends across different
environments and further refine the modelling of the UV upturn population. In the next section we describe the data and our analysis. The results are presented and discussed in section 3, while section 4 presents our conclusions. We adopt the latest cosmological parameters from \cite{PlanckCollaborationVI}. All magnitudes quoted are in the AB system.

\section{Data and Photometry}

Our dataset consists of luminous red galaxies (LRGs)
with $i < 19.8$, colour selected from the SDSS DR1 
\citep{Abazajian2003} according to the prescriptions
of \cite{Eisenstein2001} and for which new spectroscopy
was obtained within the 2SLAQ survey \citep{Cannon2006}
using the 2 degree field spectrograph \citep{Lewis2002}
on the 3.9m Anglo-Australian Telescope. The
targeted galaxies span the redshift range $0.3 < z < 0.8$ and the sample is approximately 90\% complete.
Within this sample of 15,000 galaxies we have selected for our analysis only those objects where the best
fitting spectroscopic template is the SDSS LRG template
of \cite{Eisenstein2003}, which are most likely to be
fully quiescent ETGs showing (if they resemble their
counterparts in clusters at these redshift) the
ultraviolet upturn. Inspection of the SDSS images
confirms the nature of these objects as typically
E/S0 galaxies. One caveat is 
that these field LRGs are likely to dwell within environments significantly less rich than clusters
and more typical of the general field, such as small and poor groups \citep{Tal2012,Tal2013}.

\begin{figure}
	\includegraphics[width=\columnwidth]{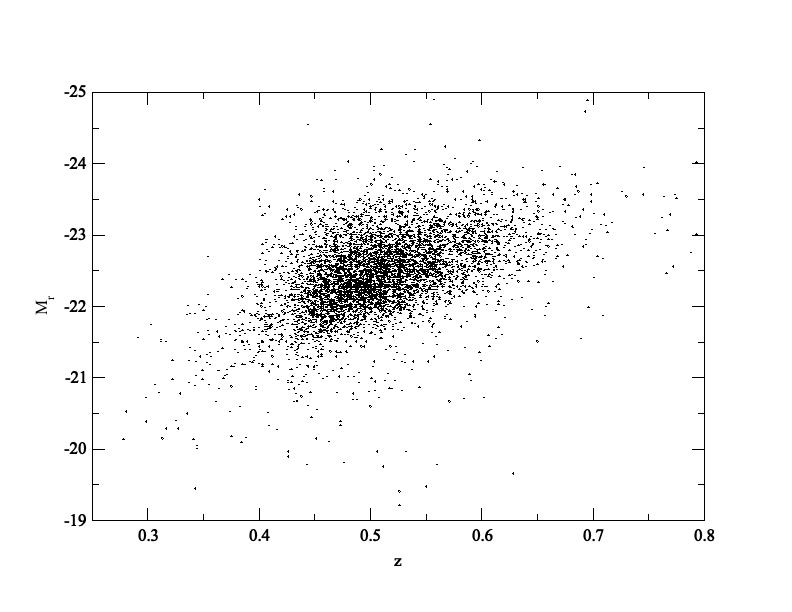}
    \caption{The magnitude and redshift distribution of the LRG sample used in this paper}
    \label{fig:sample}
\end{figure}

We have then selected those galaxies having
observations in the GALEX archive
\citep{Morrissey2005,Martin2005,Morrissey2007};
we restricted our search to objects with exposure
times in the $NUV$ band (corresponding approximately
to the rest-frame $FUV$ at the redshifts of our sample)
$> 1000$s and within $39.6'$ of the center of each of
the GALEX circular tiles, to avoid the noisier region
around the rim of each field. This leaves a total of
6094 galaxies in our sample. In Fig.~\ref{fig:sample}
we plot the absolute $M_r$ magnitude (calculated using
the conventional cosmological parameters and with
$k+e$ corrections from a \citealt{Conroy2009,Conroy2010} model 
with formation redshift of 4, solar metallicity and
star-formation e-folding time of 0.3 Gyr, that provides
a good match to the observed colours of LRGs) vs.
redshift. The rest-frame $r$ band is a good match to
the observed $i$ band for these objects (in which they were selected by the 2SLAQ survey). The $FUV$
data are less abundant and probe a much bluer wavelength regime than is conventionally used to study the ultraviolet
upturn and hence not used here.

\begin{table*}
	\centering
	\caption{Photometry for galaxies in the sample. The sample is split into cells (labelled A-R) centered on absolute magnitude $M_r$ and redshift $z$ with a given width in redshift $\delta z$ resulting in $N_{gal}$ galaxies per bin. The photometry for the resulting median stack is given in $NUV$ with its $1\sigma$ uncertainty (err) and in $i$ (from SDSS photometry), with uncertainty on the median $i$ $0.01<$}
	\label{tab:data}
	\begin{tabular}{lcccccccc} 
		\hline
ID & $M_r$ & $z$ & $\delta z$ & N$_{gal}$ & $NUV$ & err & $i$ & $NUV-i$\\
		\hline
 
A & $-21.75$ & 0.35 & 1.0 & 55 &  24.89   & 0.88  &  19.36 & 5.53 \\
B & $-21.75$ & 0.425 & 0.50 & 228 & 25.13 & 0.64  &  19.69 & 5.44 \\
C & $-21.75$ & 0.475 & 0.50 & 378 & 25.61 & 0.66  &  19.91 & 5.70 \\
D & $-22.25$ & 0.425 & 0.50 & 228 & 26.92 & 0.94  &  19.17 & 7.75 \\
E & $-22.25$ & 0.4625 & 0.25 & 412 & 25.18 &  0.40  &  19.05  &  6.13 \\
F & $-22.25$ & 0.4875 & 0.25 & 533 & 26.06 &  0.52  &  19.63  &  6.43 \\
G & $-22.25$ & 0.5125 & 0.25 & 431 & 25.05 &  0.34  & 19.70   &   5.35\\
H & $-22.25$ & 0.5375 & 0.25 & 282 & 26.05 &   0.81  &   19.83 &     6.22\\
I & $-22.75$ & 0.45   & 1.0  & 640 & 25.26  & 0.33  &   19.09  &    6.17\\
J & $-22.75$ & 0.525  & 0.50 & 725 & 25.76 & 0.43  & 19.38   &   6.38\\
K & $-22.75$ & 0.575  & 0.50 & 434 & 25.82  &  0.79 &  19.61  &    6.21\\
L & $-22.75$ & 0.625  & 0.50 & 157 & 25.13 &  0.75  &  19.78 &     5.35\\
M & $-23.25$ & 0.45   & 1.0  & 198 & 24.23 &  0.24 &  18.62 &     5.61\\
N & $-23.25$ & 0.525  & 0.50 & 212 & 24.49 & 0.34  & 18.90  &    5.59\\
O & $-23.25$ & 0.575  & 0.50 & 199 & 24.63 & 0.32 &  19.15  &    5.48\\
P & $-23.25$ & 0.65 & 1.0 & 139 & 24.97 & 0.85 & 19.46 & 5.43 \\
Q & $-23.75$ & 0.55   &  1.0 &  56 & 24.37 & 0.75 &  18.70  &    5.67\\
R & $-23.75$ & 0.65   &  1.0 &  44 & 24.36 & 0.62 &  19.04 &     5.32\\
		\hline
	\end{tabular}
\end{table*}

\begin{figure*}
\begin{tabular}{cccc}
A & B & C & D \\
\includegraphics[width=4cm]{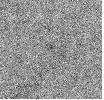} &
\includegraphics[width=4cm]{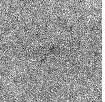} &
\includegraphics[width=4cm]{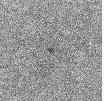} &
\includegraphics[width=4cm]{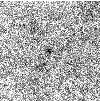} \\
E & F & G & H \\
\includegraphics[width=4cm]{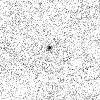} &
\includegraphics[width=4cm]{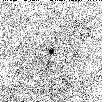} &
\includegraphics[width=4cm]{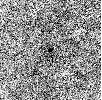} &
\includegraphics[width=4cm]{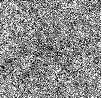} \\
I & J & K & L \\
\includegraphics[width=4cm]{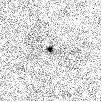} &
\includegraphics[width=4cm]{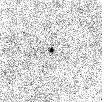} &
\includegraphics[width=4cm]{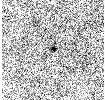} &
\includegraphics[width=4cm]{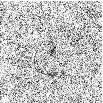} \\
M & N & O & P \\
\includegraphics[width=4cm]{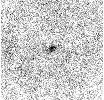} &
\includegraphics[width=4cm]{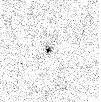} &
\includegraphics[width=4cm]{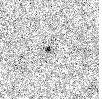} &
\includegraphics[width=4cm]{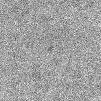} \\
Q & R & & \\
\includegraphics[width=4cm]{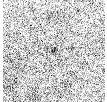} &\includegraphics[width=4cm]{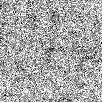} & &  \\
\end{tabular}
\caption{$NUV$ images of the median stacks for 2SLAQ LRGs in the magnitude/redshift cells identified in Table~\ref{tab:data}. The letters above each postage
stamp are the identifier in the table}
\label{fig:med}
\end{figure*}

We isolated a $150'' \times 150''$ postage stamp in the GALEX image around the optical position of each galaxy. Where more than one observation was available we averaged all images to produce a single image for 
each galaxy. As expected, given the sensitivity of GALEX, only a few objects are
individually weakly detected in $NUV$, even in total exposures of
several hundred ks. Consequently we stacked the data using the following scheme. We divided our sample into 
several cells in absolute magnitude and redshift,
each containing typically a few hundred galaxies
(except for a few low luminosity bins at low redshift
and the two highest luminosity bins at high redshift),
and produced a median stack for all GALEX postage
stamps for each cell. By using the median of typically several hundred galaxies, we can remove the
influence of outliers (e.g., objects with star formation) from our photometry.  These
images are shown in Fig.~\ref{fig:med}
where the resulting objects are generally well
detected.

At the typical redshifts of these galaxies, and given
the GALEX pixel scale of $1.5''$ and point spread function of $4.5''$ our objects are essentially unresolved in the $NUV$ data and we therefore carry out photometry within
a $10''$ radius aperture, that would include $>97\%$
of the light for a star.

We followed the procedure by \cite{Atlee2009} and
calculated the uncertainties on the mean magnitude by bootstrap re-sampling.
This includes both counting statistics and the 
effects of intrinsic scatter. The bootstrapping
analysis folds all intrinsic variations in the sample into the uncertainty on the mean magnitude. We drew 100
bootstrapping realizations in each magnitude/redshift bin and used the RMS of the resulting magnitudes as the uncertainty on the mean magnitude. 

We determined the mean foreground extinction for our galaxies
using the values by \cite{Schlafly2011} and the 
Milky Way extinction law of \cite{Cardelli1989}: 
these are 0.29 mag. for the $NUV$ and 0.07 mag. for
the $i$ band. Internal extinction is ignored as LRGs are typically devoid of dust and gas. Observations of local SDSS LRGs by \cite{Barber2007} returned a mean extinction of $A_z=0$, while even for cluster early type galaxies at $z=1.3$ \cite{Rettura2006} measured internal 
extinctions $E(B-V) < 0.05$.

\begin{figure*}
\includegraphics[width=0.45\textwidth]{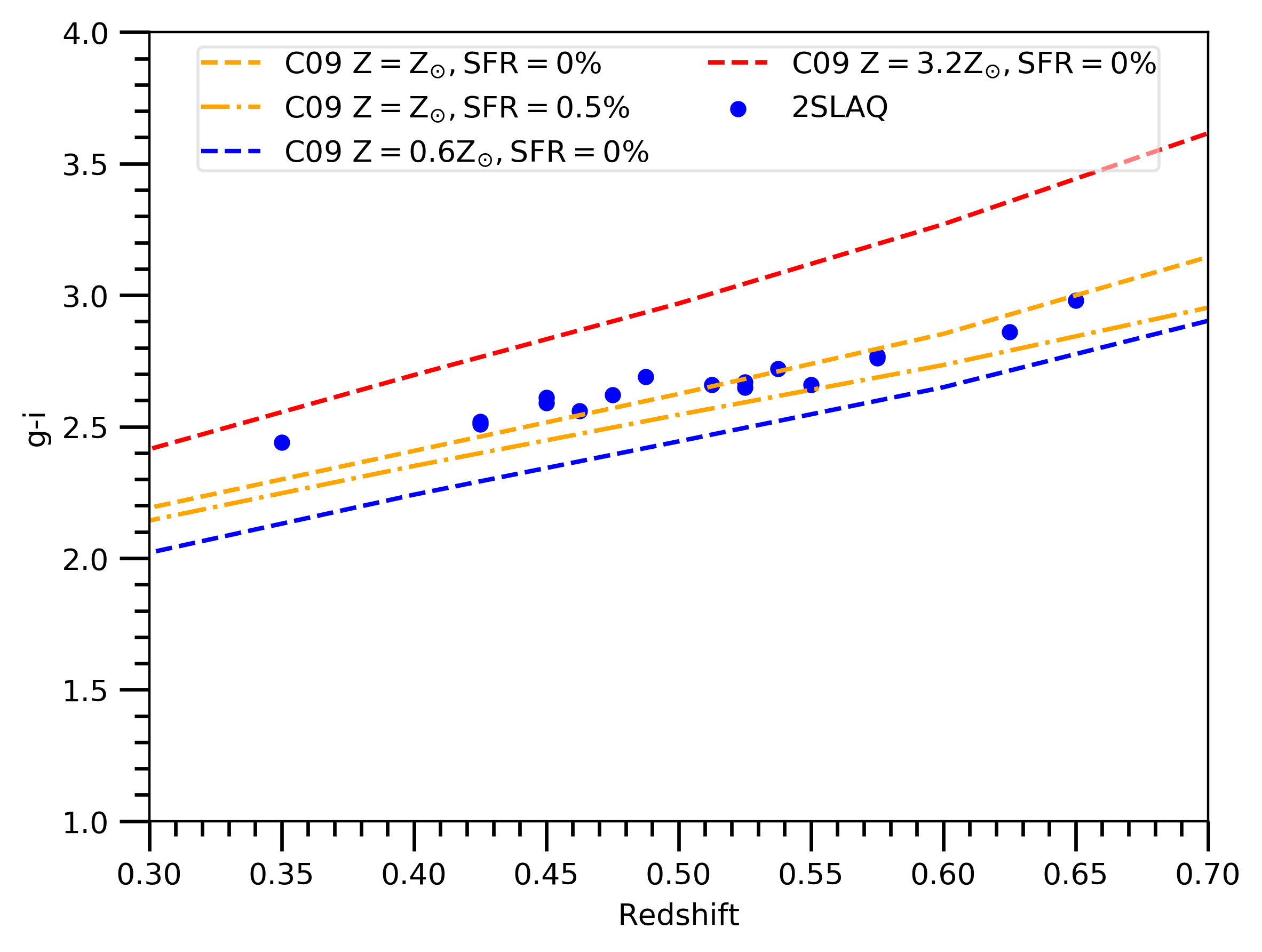}
\includegraphics[width=0.45\textwidth]{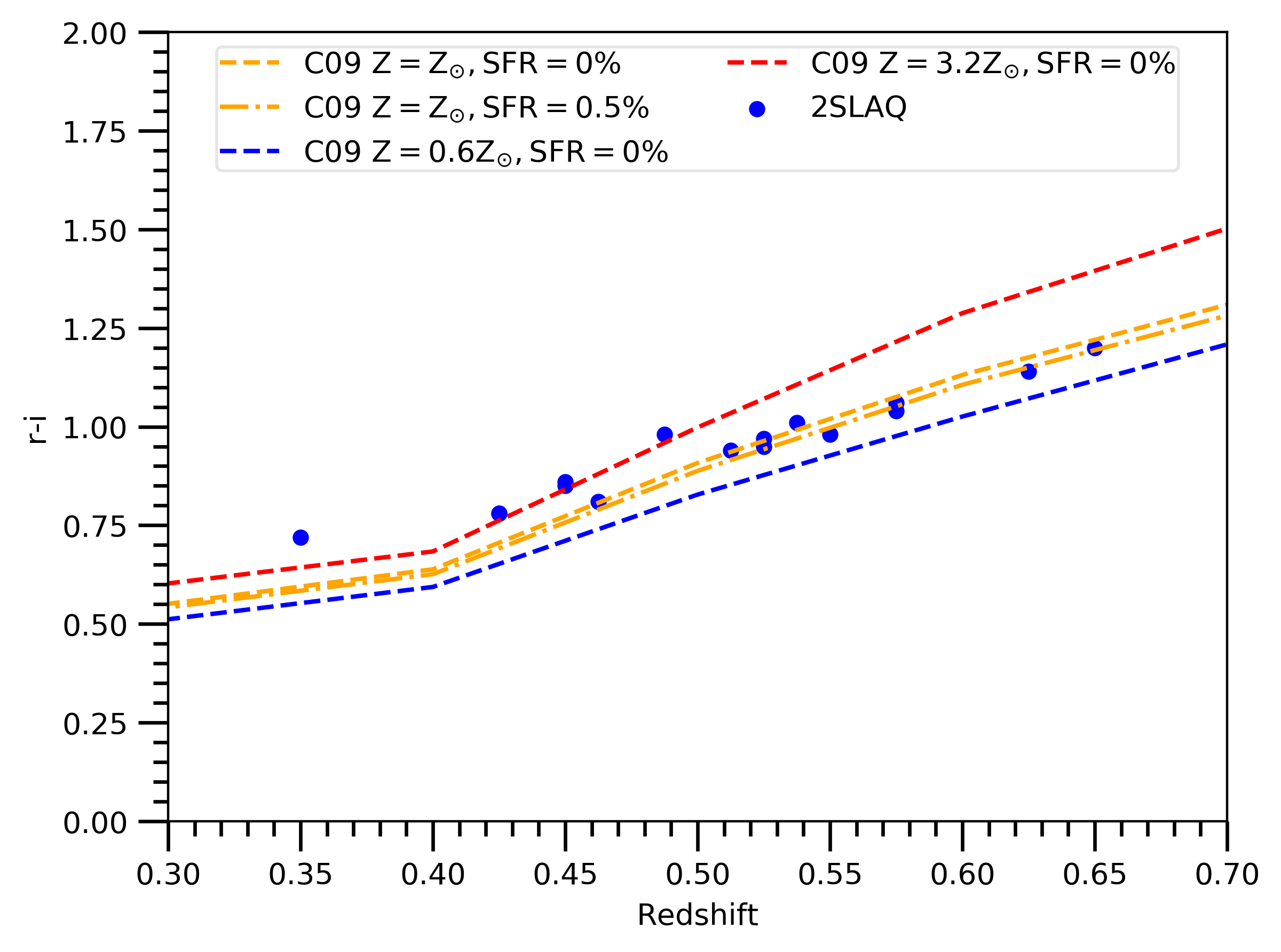}
\caption{Comparison of observed $g-i$ and $r-i$ median colours for 2SLAQ galaxies in this sample (see Table 1) with models from Conroy et al. (2009); Conroy \& Gunn (2010) with formation redshift of 4, $\tau=0.3$ Gyr and various metallicities. This also includes a comparison with a model with residual star formation forming 0.5\% of the stellar mass at a constant rate after the initial burst.}
\label{fig:colors}
\end{figure*}

In Fig.~\ref{fig:colors} we plot the observed $g-i$ and $r-i$ median colours for galaxies in these stacks together with several models from \cite{Conroy2009,Conroy2010}. The solar metallicity model with $z_f=4$ and $\tau=0.3$ Gyr provides a very good 
representation of these data, with any scatter in the colours being explained by a small spread in age/metallicity, confirming
that the rest frame optical colours of our galaxies show no evidence of current star formation. The stacked optical data strongly imply that only a subset of the passively evolving models, namely those with solar metallicity (including  \citealt{Bruzual2003} and \citealt{Conroy2009}) and reject significantly higher and lower metallicity models at all redshifts probed here.

\section{Discussion}

\subsection{Comparison with models}

\begin{figure*}
    \centering
    \includegraphics[width=\textwidth]{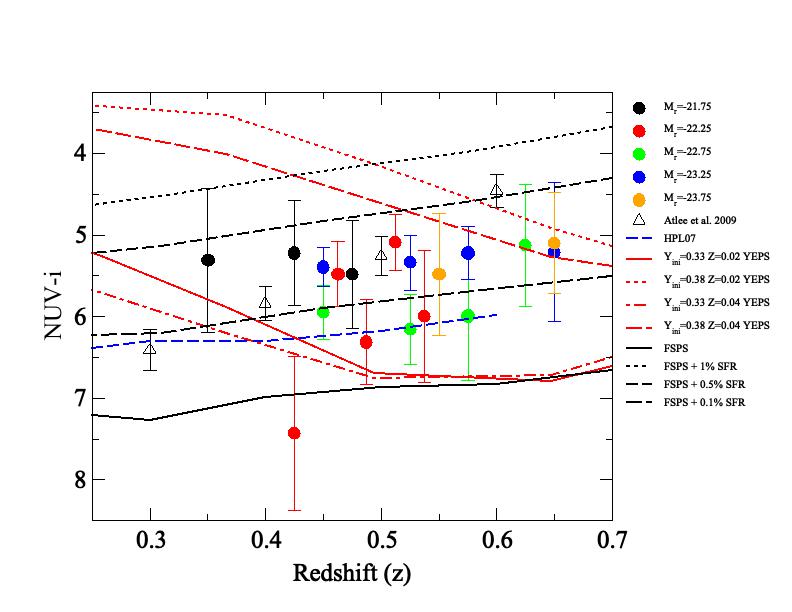}
    \caption{$NUV-i$ colors for 2SLAQ galaxies as a function of redshift and absolute $r$ magnitude ($k+e$ corrected as described in the text): see legend in the figure. We also show data from Atlee et al. (2009) in the same redshift range.}
    \label{fig:comp1}
\end{figure*}

Fig~\ref{fig:comp1} compares our colours for LRGs in the 2SLAQ sample with those from \cite{Atlee2009}, where we used a model from \cite{Conroy2009} with solar metallicity, formation redshift of 4 and star formation e-folding time of 0.3 Gyrs (values that typically reproduce ETG colors at all redshifts below 1.5) to convert from their $NUV-V$ to our $NUV-i$\footnote{As in previous work we adopt the models by \cite{Conroy2009} as these make no attempt to semi-empirically fit the UV upturns as do \cite{Bruzual2003}}. There
is reasonably good agreement, given the differences in selection, between our data and 
those of \cite{Atlee2009} except for their
highest redshift point that appears to be
too blue (this may be due to the small number of such galaxies in their sample and selection effects noted by these authors). We also plot a model from \cite{Conroy2009} with formation redshift of 4, solar metallicity and star formation e-folding time of 0.3 Gyr (same as we used above) to show how this does not 
reproduce the observed $NUV-i$ colours of
field LRGs: these are typically 1.5--2 mag. bluer, the typical excess flux of the UV upturn (recall that these colours are nearly equivalent to rest-frame $FUV-r$). We also took a single stellar population from \cite{Han2007} and converted their colours
to our observed $NUV-i$, using the $k-$corrections from \cite{Atlee2009}. While these colours are bluer than the model by \cite{Conroy2009} and \cite{Conroy2010} which explicitly avoid modelling the UV upturn, they do not match the observed colours of 2SLAQ LRGs, being about 1 mag. too red. Indeed, it has already been argued that the sdB stars in our Galaxy are redder (in UV colours) than the UV upturn in ETGs \citep{Yi2008}.

However, field LRGs may be less quenched than cluster
ETGs and show residual star formation at low levels. \cite{Rusinol2019} use UV spectroscopic indices 
to claim that their sample of BOSS LRGs at $z=0.4$ shows residual star formation at about the 1\% level, although the data may also be interpreted as being produced by blue HB stars. To test this, we have added a constant rate of star formation to the original model of \cite{Conroy2009} and \cite{Conroy2010}, with a mass fraction formed during a constant mode of star formation of 1\%, 0.5\% and 0.1\% of the total mass, added to the initial burst at high redshift.  The observed $NUV-i$ colours may be qualitatively explained by models forming
between 0.1 and 0.5\% of the total mass in a regime of
constant star formation superposed over the original
burst at $z=4$. We also show the effect of this constant rate of star formation in Fig.~\ref{fig:colors} on the optical colours, showing that they are essentially unchanged from those of models with no extra star forming component. However, these models become about 1 mag. bluer across the redshift range we explore, whereas we see no evidence of such evolution in the observed colours over this redshift range. 

The star formation rates of local ellipticals can
be estimated from the rate of core collapse supernovae.
For $z <0.2$ ETGs in the SDSS, \cite{Sedgwick2021} estimate star formation rates of approximately 0.1 $M_{\odot}$ yr$^{-1}$, although \cite{Irani2021} 
argues that these are likely to be overestimated
by up to an order of magnitude owing to the mis-identification of core collapse supernovae
and contamination from supernovae associated with
nearby diffuse galaxies. This latter is of course a possible source of contamination in our data as well.
These star formation rates are equivalent, if they
are constant, to one of the \cite{Conroy2009,Conroy2010}
models forming 0.01\% of the total mass since $z=4$.
Star formation rates of this magnitude cannot therefore
be excluded, based on our data alone, although they are lower than those inferred by \cite{Rusinol2019}. However, resolved
images of nearby galaxies do not show any of the expected populations of B stars or earlier type that would be needed \citep{OConnell1992}, while $FUV$ 
images of nearby ellipticals from {\it Astro HUT} 
show none of the clumpiness  otherwise associated 
with star-forming regions elsewhere. Nevertheless, 
only observations at higher ($z \sim 1$) can 
truly discriminate a UV upturn component from other models.

In this figure we also compare
our data with models from \cite{Chung2011,Chung2017}. These include a population of stars with extra helium enrichment and have been shown to account
for the evolution of the UV upturn colours in cluster ETGs \citep{Ali2018a,Ali2018b,Ali2018c}. These models assume $z_f=4$, metallicity of 0.02 and 0.04, and $Y_{ini}=0.23$ (cosmological value), 0.28, 0.33 and 0.38 for a chemical infall model (where $Y=Y_{ini} + 2Z$). We do not plot the two models with lower He abundance as they lie very close to the standard FSPS model with no upturn, to avoid confusion. The data are not compatible with models of standard (i.e., cosmological He abundance) composition, as shown in Fig.~\ref{fig:comp1} above as well, but are well encompassed by models with $0.33 < Y_{ini} < 0.38$ (i.e., $Y=0.37 - 0.42$ assuming solar metallicity -- Fig. 3). There are no detectable effect on optical colours. This is similar to what is observed in clusters of galaxies, although the required helium abundance is slightly lower than inferred in \cite{Ali2018c}. This may be partly explained by differences in the ages and star formation histories of 
ETGs in clusters and the field, with field galaxies expected to be somewhat younger and to have had a more extended star formation history. 

However, unlike the work in clusters by \cite{Ali2018c} and \cite{Ali2021}, we do not here
reach the redshift where the UV upturn 
colours start to evolve strongly, which is at $z > 0.7$ in galaxy clusters, where we may more strongly discriminate between models. While we do not see (or expect to see) this strong evolution in colour given the redshift range probed, the strong similarity in the behaviour of this sample and that of ETGs in clusters, indicates that the same process is likely to be responsible for the UV flux in both environments. Given that the best explanation for
the behaviour displayed by cluster ETGs is the presence of a He-rich population, ``switching on'' the blue HB at $z=0.7$, the simplest explanation of the result shown here is the same. Clearly a test of this will be attempting to explore the strength of the UV upturn in field galaxies at higher redshift, the subject of a future paper.

\subsection{Variations with metallicity and age}
In this paper we adopt a single fiducial model for age, metallicity and star formation e-folding time over which a He-rich subpopulation, with various degrees of He enrichment, or a binary population or residual constant star formation are superposed. 

The colour of the UV upturn in the models of \cite{Conroy2009,Conroy2010} and \cite{Chung2011,Chung2017} does not change strongly with age, as the horizontal branch stars do not evolve beyond the RR Lyrae gap except for ages far in excess of the 
Hubble time. It is only He, or residual star formation,
respectively, that can drive these models' UV colours to
the blue. 

Low metallicity populations for these models can be
excluded by the poor match to the optical colours that
would result (see Fig.~\ref{fig:comp1}) and by the
previous observation that a significant contribution
by low metallicity stars would strongly weaken the
CN and Mg lines observed in the spectra of LRGs. 
If residual star formation takes place in more metal poor gas, this would of course also result in much bluer optical colours than we observe.

\cite{Han2007} do not consider lower metallicity stellar populations for their binary model. We would expect such stars to be bluer purely by the effects of line blanketing; however, the same consideration applies as
to the effect of such a low metallicity population on the otherwise strong metal lines in the ultraviolet spectra of LRGs.

\subsection{Interpretation}

These observations indicate that the UV upturn is a fundamental property of the stellar populations of ETGs across all environments \citep{Ali2019,Phillipps2020} from field to cluster, irrespective of environment at
least below $z=0.7$. Since field LRGs are likely
to be less quenched than cluster ETGs it is appropriate to consider the level of residual star formation within them. Given that we see that the 
same level of UV emission in both field and cluster ellipticals and we know that the UV emission in cluster galaxies does nor arise from star formation,
these observations argue against a significant role
for star formation in the production of the UV
flux observed in these galaxies. One caveat is 
that these field LRGs are likely to dwell within relatively massive halos, and therefore sample a comparatively denser environment \citep{Tal2012,Tal2013}.

These data on their own do not suffice to 
establish the cause of the UV upturn in the field population, as they do  not yet reach the putative redshifts at which strong evolution in this colour 
is expected to occur due to the blue HB being populated several Gyr after the zero-age main sequence \citep{Chung2011,Chung2017}. Nevertheless, the similarity between field and cluster galaxies,
and the similar lack of evolution below $z=0.6$ in the LRG sample of \cite{LeCras2016} tends to
suggest that a similar phenomenon is being 
observed across a wide range of environmental
densities. It appears likely that the UV upturn represents the occurrence of a novel enrichment 
channel taking place within 2 Gyrs of the Big Bang across a wide variety of environments and over several orders of magnitude in stellar mass to produce a sub-population with excess He. However, these current observations do not give an indication as to the nature of this channel. This is likely to require spectroscopic observations at significantly higher redshift. Despite this, we know that such He rich stars exist within globular clusters and other systems. Identifying the enrichment pattern responsible will likely solve both problems.

\section*{Acknowledgements}

The 2SLAQ surveys were conducted by a substantial team of people in the UK and Australia.

Funding for the creation and distribution of the SDSS archive have been provided by the Alfred P. Sloan Foundation, the Participating Institutions, NASA, NSF, the US Department of Energy, the Japanese Monbukagakusho and the Max Planck Society. The SDSS Web site is http://www.sdss.org.

This research is based on observations made with GALEX, obtained from the MAST data archive at the Space Telescope Science Institute, which is operated by the Association of Universities for Research in Astronomy, Inc., under NASA contract NAS 5–26555.

C.C. acknowledges the support provided by the National Research Foundation of Korea (2022R1A2C3002992).

\section*{Data Availability}

Data can be retrieved from the relevant archives (SDSS, 2SLAQ) as mentioned in the text.



\bibliographystyle{mnras}
\bibliography{example} 




\appendix


\bsp	
\label{lastpage}
\end{document}